\documentclass[12pt]{article}

\usepackage{amsmath}
\usepackage{graphicx}
\usepackage{color}
\usepackage[compat=1.0.0]{tikz-feynman}
\usepackage{feynmp-auto}
\usepackage[colorlinks=true, urlcolor=blue, linkcolor=blue, citecolor=blue]{hyperref}
\usepackage{cite}

\def \lsim{\mathrel{\vcenter
     {\hbox{$<$}\nointerlineskip\hbox{$\sim$}}}}

\newcommand{\beq}{\begin{equation}}
\newcommand{\eeq}{\end{equation}}
\newcommand{\beqa}{\begin{eqnarray}}
\newcommand{\eeqa}{\end{eqnarray}}  
\newcommand{\beqar}{\begin{eqnarray*}}
\newcommand{\eeqar}{\end{eqnarray*}}

\begin{document}
\thispagestyle{empty}

$\,$

\vspace{-10pt}

\begin{center}

\textbf{\Large High energy gamma rays and neutrinos from\\
the Sun, Jupiter and  Earth}

\vspace{50pt}
Pablo de la Torre, Miguel Guti\'errez, Manuel Masip, Alejandro Oliver
\vspace{16pt}

\textit{Departamento de F{\'\i}sica Te\'orica y del Cosmos}\\
\textit{Universidad de Granada, E-18071 Granada, Spain}\\
\vspace{16pt}

\texttt{pdelatorre@ugr.es, mgg@ugr.es, masip@ugr.es, alejandroliver@correo.ugr.es}

\end{center}

\vspace{30pt}

\date{\today}

\begin{abstract}

Cosmic rays reaching the atmosphere of an astrophysical object produce showers of 
secondary particles that may then escape into space. 
Here we obtain the flux of gamma rays and neutrinos 
of energy $E>10$ GeV emitted by the Sun, Jupiter and Earth. 
We show that, while the solar magnetic
field induces a flux of gamma rays from all the points on the Sun's surface, the dipolar magnetic field 
in the  planets implies  high energy photons only from the very peripheral region. Neutrinos, in contrast, can cross 
these objects and emerge from any point on their surface. 
The emission from these astrophysical objects exceeds the diffuse flux from cosmic ray interactions with the interstellar medium and has a distinct spectrum and gamma ray to neutrino ratio.

\end{abstract}

\vfill
\eject

\section{Introduction} 

Cosmic rays (CRs) accelerated in supernova remnants are trapped by Galactic magnetic fields  for  around a million years before they escape into intergalactic space. During this time a small fraction of them collide with the interestelar gas (hydrogen and helium in a proportion of 3 to 1 in mass), producing a diffuse flux of secondary gamma rays and neutrinos \cite{Strong:1998fr} that reaches the Earth.
This diffuse  flux is proportional to the column density of gas along the line of sight modulated by the CR density:  the effective depth varies between 
0.3 g/cm$^2$ towards the center  of the galaxy and 0.002 g/cm$^2$ in the opposite (anticenter) direction \cite{Carceller:2016upo}. At  $10$--$10^6$ GeV the  gamma and neutrino fluxes are of similar intensity and concentrated around the Galactic equator (50\%  from $|b|\le 5^\circ$), with a spectrum ($\approx E^{-2.6}$) that is similar to the one in the parent CR flux \cite{Lipari:2018gzn,Schwefer:2022zly}.
In addition, the Galactic diffuse fluxes also include a contribution from the ensemble of unresolved sources in our galaxy. For gamma rays this contribution is expected to be dominated by pulsars, which produce leptonic gamma rays through $e^{\pm}$ bremsstrahlung and inverse Compton scattering.

We will focus here on another Galactic source of gamma rays and neutrinos: astrophysical objects like the Sun \cite{Seckel:1991ffa} and planets with a thin atmosphere like Jupiter or  Earth. These compact objects absorb high energy CRs and {\it process} a fraction of their energy into gamma rays and neutrinos that  are then emitted to  space.
Our interest is twofold. If the object is resolved, detecting these fluxes would provide information about its magnetic field, its surface and its interior.  
If unresolved, these emissions will also give a small contribution with a distinct spectrum and
$\gamma$ to $\nu$ ratio to the Galactic diffuse fluxes. At any rate, these high energy fluxes are an irreducible background in indirect dark matter searches \cite{PerezdelosHeros:2020qyt,Linden:2024uph}. Our objective is then to understand and characterize the high-energy photon  and neutrino fluxes emitted by each type of object. 

Generically, compact objects will absorb gamma rays while letting most neutrinos escape. This will be the case when the dominant magnetic field is dipolar, as in the case of Earth or Jupiter, where we may only observe gamma rays from the very peripheral regions. However this is not the case in objects with open magnetic field lines and convection (produced by the emission of plasma) near the surface. At 10--1000 GeV these {\it albedo} photons from CRs have been observed by Fermi-LAT \cite{Fermi-LAT:2011nwz,Linden:2018exo} and HAWC \cite{HAWC:2022khj}, while the neutrino flux is currently being searched for at telescopes like IceCube \cite{IceCube:2019ubb} and KM3NeT \cite{Gutierrez:2023aaf}. In any case and in contrast to the hadronic emission from CR collisions with interstellar gas, the emission from compact objects strongly discriminates between gamma rays and neutrinos.

\begin{figure}[!t]
\begin{center}
\includegraphics[scale=0.4]{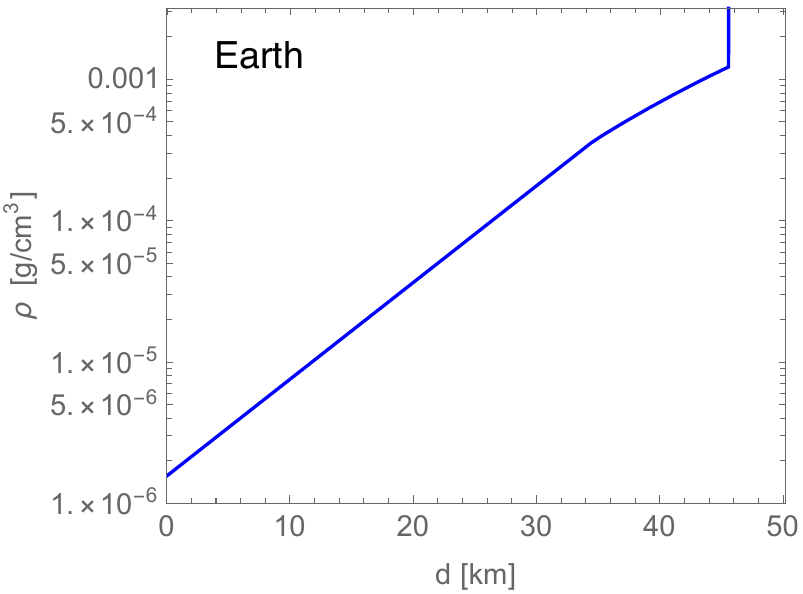}\hspace{0.1cm}
\includegraphics[scale=0.4]{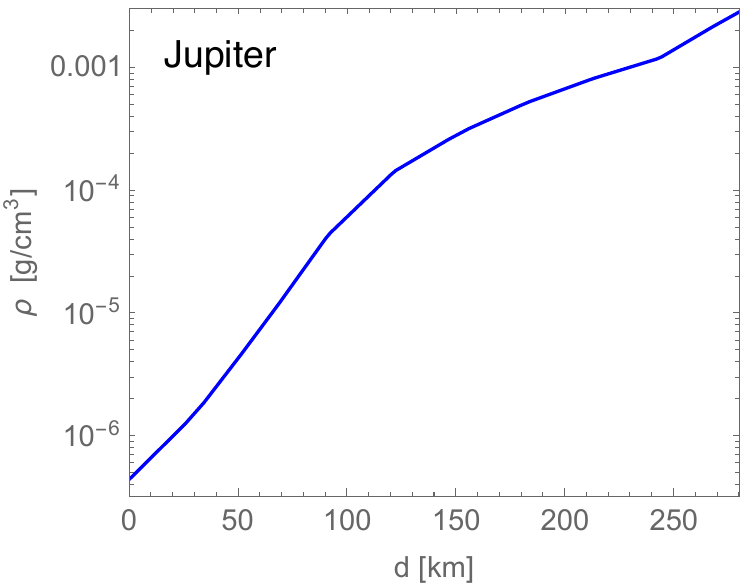}\hspace{0.1cm}
\includegraphics[scale=0.4]{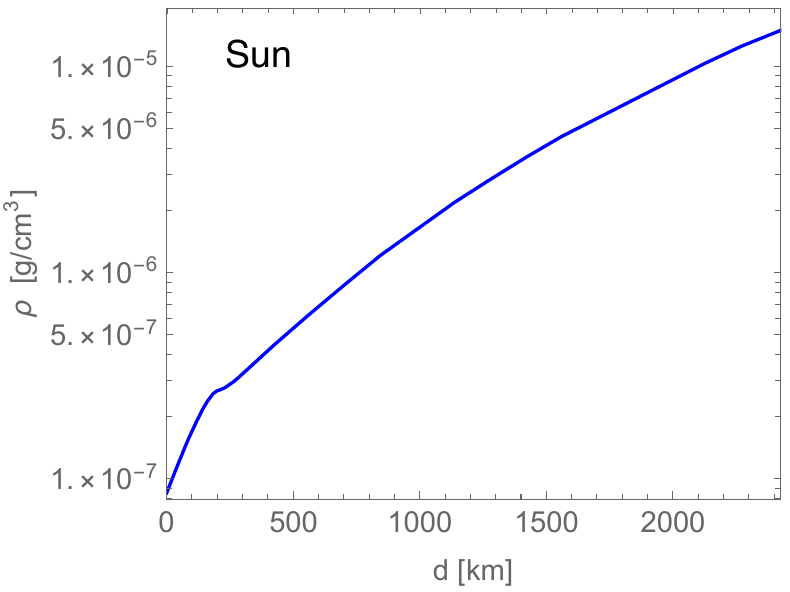}
\end{center}
\vspace{-0.3cm}
\caption{Matter density encountered by a particle approaching vertically the Earth, Jupiter and the 
Sun after crossing a distance $d$ 
(initial point at a depth of 1 g/cm$^2$ and final point at 1000 g/cm$^2$). 
\label{f1}}
\end{figure}

The three astrophysical objects under study have very different size, density profile and magnetic field, but they share a low mass density near the surface: they have an atmosphere that allows for a significant fraction of the high energy mesons in a CR shower decay and produce leptons. This is in contrast to what happens when a CR hits the surface of the Moon: all its energy will be deposited within 10 meters of rock, producing only very low energy neutrinos ($E<100$ MeV). 
In Fig.~\ref{f1} we plot the density encountered by a CR moving vertically towards the surface of each object from an initial point at a depth of 1 g/cm$^2$ down to a depth of 1000 g/cm$^2$, where the hadronic component of the shower has been completely absorbed. To obtain these lines we have used the solar model in \cite{Christensen-Dalsgaard:1996hpz} and the atmospheric models in  \cite{Lipari:1993hd} for Earth and \cite{Seiff:1998} for Jupiter. We find that it takes around 46 km in the Earth, 280 km in Jupiter or 2400 km in the Sun to cross a column density of 1000 g/cm$^2$.

In our analysis we will take a primary flux  \cite{Boezio:2012rr} made of protons and He nuclei; 
at energies (per nucleus) below
 $E_{\rm knee}=10^{6.5}$ GeV
\beq
\Phi_p = 1.3 \left( {E\over {\rm GeV}} \right)^{-2.7}{\rm (GeV\,cm^2\,sr\,s)^{-1}};
\hspace{0.5cm}
\Phi_{\rm He} = 0.54 \left( {E\over {\rm GeV}} \right)^{-2.6} {\rm (GeV\,cm^2\,sr\,s)^{-1}},
\label{fluxp}
\eeq
whereas at  $E>E_{\rm knee}$ we assume continuity and a common spectral index $\alpha=-3$.

\section{High energy particles from the Sun} 

Although the solar emission induced by CRs was already discussed more than 30 years ago \cite{Seckel:1991ffa}, 
a precise calculation is in principle plagued by the uncertainties introduced by the solar magnetism \cite{Orlando:2020ezh,Mazziotta:2020uey,Li:2023twp,Li:2025xxi}. 
Here we will use the framework proposed in \cite{Gutierrez:2022mor} slightly modified to incorporate the results
on the solar gamma flux at 1 TeV recently obtained by HAWC \cite{HAWC:2022khj}. 

The first key element to calculate this solar emission is the absorption rate of CRs at different energies. Suppose there were no
solar magnetism and CRs follow straight lines. Then the trajectories aiming to the Earth but 
absorbed by the Sun would define a black disk of radius $r\approx 0.26^\circ$, the angular size of the Sun as seen from 
the Earth. Indeed, this is what we will see
at very high energies, when the deflection of CRs by the solar magnetic field is negligible, but not at lower energies.
CRs of $E<100$ TeV are affected by a magnetic field that is
very involved. 
First of all, it has a radial component (open lines that define the Parker interplanetary field \cite{Tautz:2010vk})
that grows like $1/R^2$ as CRs approach the surface. This  gradient  may induce a magnetic mirror effect: CRs
going towards the Sun tend to bounce back. In addition, the solar wind induces convection, {\it i.e.}, CRs are propagating in a
plasma that moves away from the Sun and makes it more difficult to reach the surface. Finally, closer to the Sun the magnetic
turbulence increases and there appear  field lines that start and end on the solar surface.
Hopefully, we can understand the absorption rate of CRs with no need to solve these details,
just by using the data on the CR shadow of the Sun together with Liouville's theorem.

The data is provided by HAWC \cite{Enriquez-Rivera:2015kft}, that has studied the energy-dependence of the CR shadow during
a solar maximum. The shadow appears
at 2 TeV; it is not a black disk of $r=0.26^\circ$ but a deficit that extends into a larger angular region. By integrating it we 
find that at 2 TeV it accounts for a 6\% of a black  disk, the deficit grows to 27\% at 8 TeV, and at 50 TeV it becomes 
a 100\% deficit, {\it i.e.}, a complete black disk in the Sun's position diluted into a larger circle or $2^\circ$ radius.

\begin{figure}[!t]
\begin{center}
\includegraphics[scale=0.4]{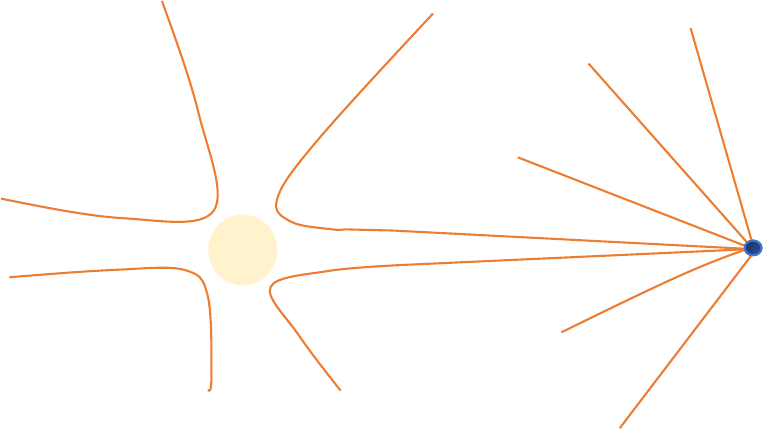}\hspace{1cm}
\includegraphics[scale=0.4]{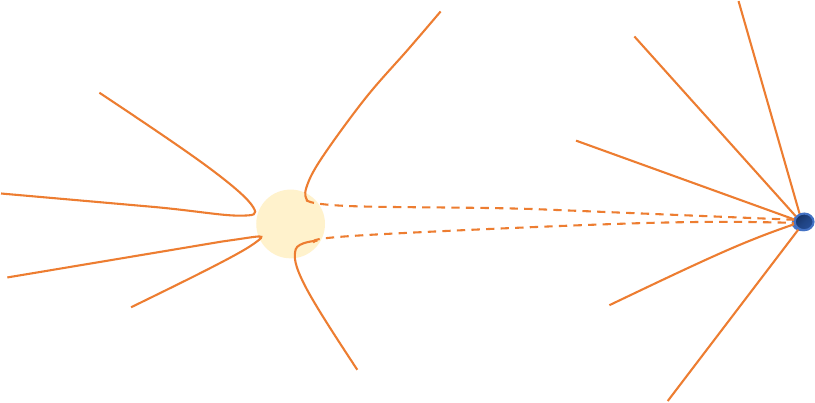}
\end{center}
\vspace{-0.3cm}
\caption{Schematic CR trajectories in the vicinity of the Sun. As the energy grows trajectories that were supposed to reach
the Earth cross a larger depth of solar matter, increasing the probability that CRs are absorbed and define a shadow.
\label{f2}}
\end{figure}

HAWC's data suggests a simple interpretation based on Liouville's theorem. 
The theorem implies that when the isotropic CR flux crosses the solar
magnetic field, it stays isotropic, and that the only possible effect of the Sun is to interrupt some of the 
trajectories that were aiming to the Earth. As we illustrate in Fig.~\ref{f2}, the solar magnetic field
deflects some of the trajectories directed to the Earth, but other trajectories will now reach us and the net
effect should be zero: an isotropic flux crossing a static magnetic lens, including a mirror, will stay 
isotropic, and the only
possible effect is to create a shadow. 
At low energies HAWC sees no shadow, meaning that a negligible fraction of the CR flux reaches the solar surface. 
At higher energies, however, 
CRs that were supposed to reach the detector hit before the Sun and are absorbed (Fig.~\ref{f2}-right). 
Therefore, studying the shadow we may deduce 
the average depth of solar matter crossed by CR's of different energy in their way to the Earth.

\begin{figure}[!t]
\begin{center}
\includegraphics[scale=0.47]{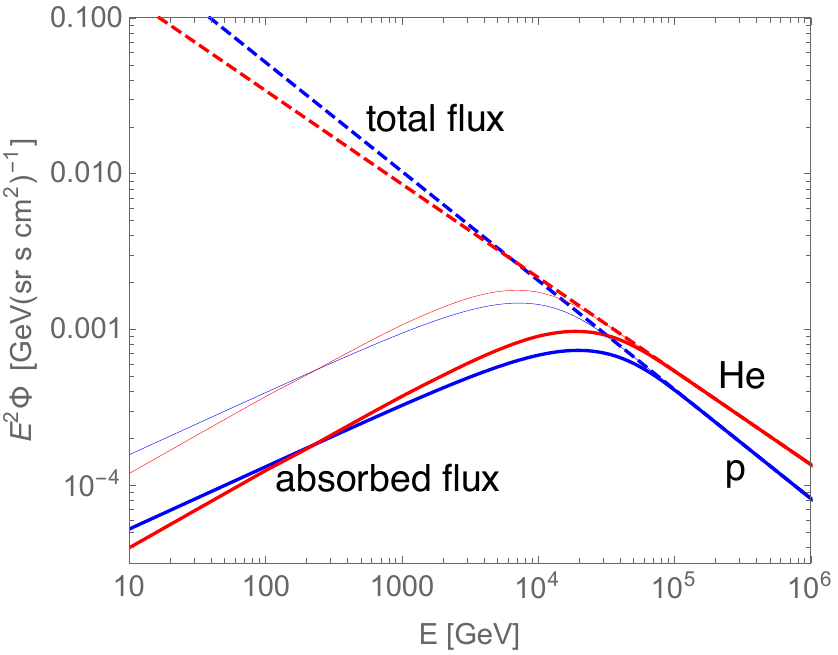}\hspace{0.5cm} 
\includegraphics[scale=0.55]{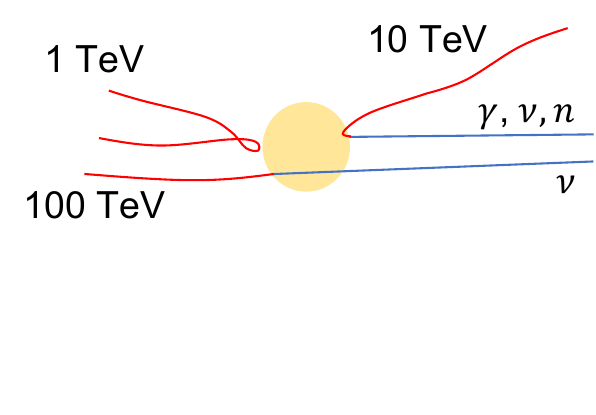}
\end{center}
\vspace{-0.3cm}
\caption{Absorbed proton and He fluxes during a solar maximum (thick) and a solar minimum (thin). On the right, 
typical CR trajectories at different energies.
\label{f3}}
\end{figure}

When a CR proton crosses an average depth  $\Delta X_{\rm H}(E)$ 
the probability that it is absorbed is
\beq
p^{\rm H}_{\rm abs}=1-\exp{\left(-{\Delta X_{\rm H}\over \lambda_{\rm int}^{\rm H}}\right)}\,.
\eeq
To explain HAWC's data we take 
\beq
{\Delta X_{\rm H}\over \lambda_{\rm int}^{\rm H}}=b_H \,E^{1.1}\,,
\label{depth}
\eeq
with $E$ expressed in GeV and a time dependent parameter $b_H$ that changes from $1.6\times 10^{-5}$ 
during a solar maximum 
to $4.8\times 10^{-5}$ during a minimum. Since the trajectory of a CR only depends on its magnetic rigidity, He nuclei 
of twice the energy will cross the same average depth and
\beq
b_{\rm He} = {b_{\rm H}\over 2^{1.1}} \,{\sigma_{{\rm He}}(E)\over  \sigma_{H}(E/2)}\,.
\eeq
Eq.~(\ref{depth}) is then the first and key hypothesis in this framework. It implies the absorption of CR primaries
given in Fig.~\ref{f3}-left. 
This absorption determines whether the CR shadow that we see at different energies is partial or complete. At low energies CRs are unable to reach the solar surface: the average depth of solar matter that they cross is small, they are not absorbed and we  see no shadow. At very high energies CRs that were supposed to reach the Earth through the solar disk find before a large column density of  matter and are absorbed, thus we see a complete shadow. Our choice for the 1.1 spectral index (we have reduced in 1\% the value used in \cite{Gutierrez:2022mor}) and the value of $b_H$ during an active phase of the Sun are based only on HAWCs observations, whereas the value of $b_H$ during a quiet Sun provides our best fit for the Fermi-LAT data (see below).
 
\begin{figure}[!t]
\begin{center}
\includegraphics[scale=0.43]{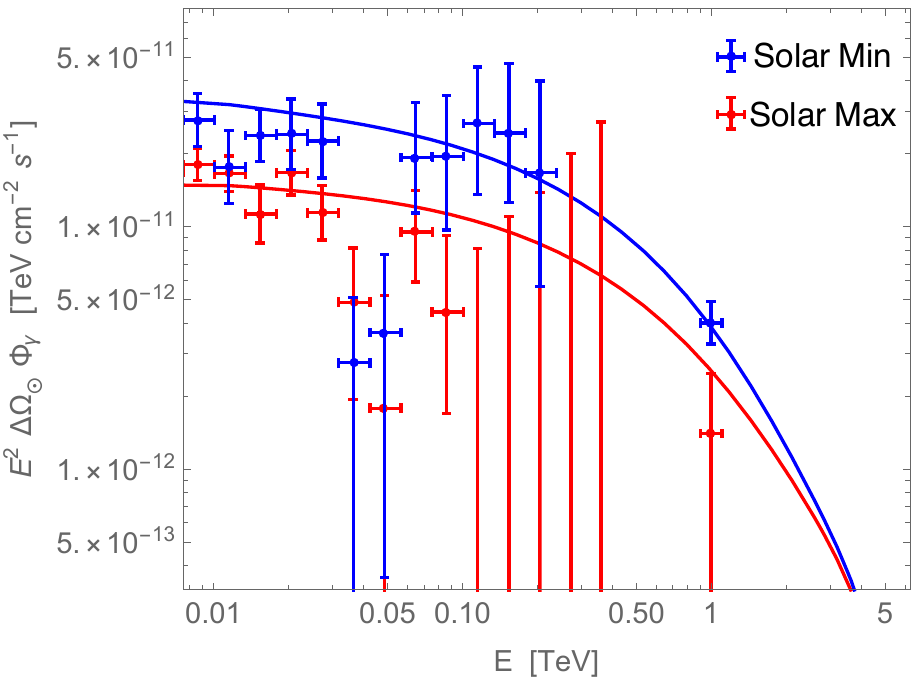}\hspace{0.5cm}
\includegraphics[scale=0.46]{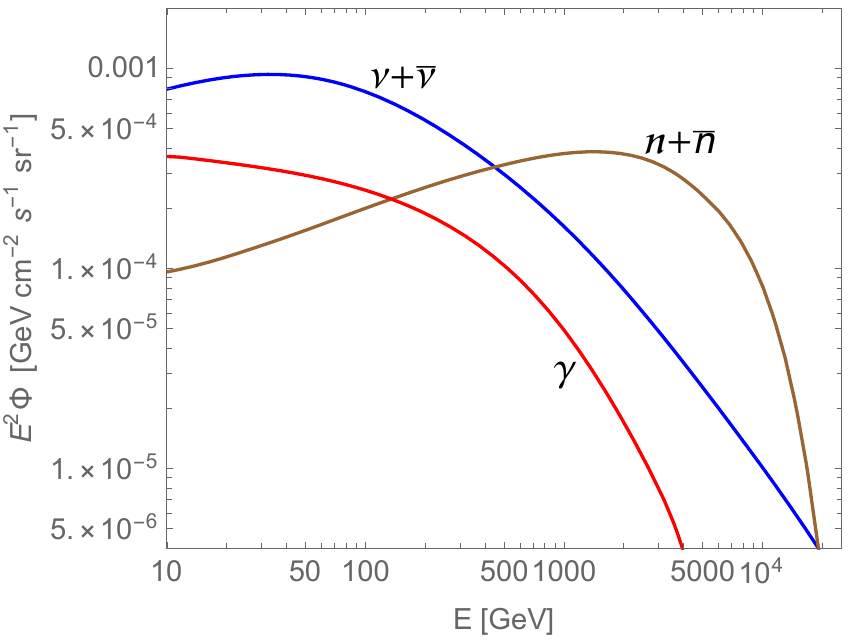}
\end{center}
\vspace{-0.3cm}
\caption{Gamma ray flux from the solar disk  (data at $E\le 200$ GeV from Fermi-LAT  \cite{Linden:2018exo}
and at $1$ TeV from HAWC \cite{HAWC:2022khj}) and our average gamma, neutron and neutrino fluxes from the Sun.
\label{f4}}
\end{figure}

Next we need to model the showering of these absorbed fluxes. A numerical simulation \cite{Masip:2017gvw}
shows that at TeV energies only trajectories very aligned with the open field lines are able to reach the Sun's
surface. Once there, CRs will shower; some of the secondaries will be emitted 
inwards, towards the Sun, but others will be emitted outwards and may eventually reach the
Earth. The probability that a secondary particle contributes to this solar albedo will depend on the direction it is emitted.
We will then assume that secondaries produced by a parent of energy $E$ above some critical 
energy $E_{\rm c}=5$ TeV (versus
3--6 TeV in \cite{Gutierrez:2022mor}) will most likely be emitted towards the Sun, whereas 
lower energy primaries will exit in a random direction:\footnote{Following \cite{Gutierrez:2022mor}, we assume that secondary charged particles of $E>E_c$ keep penetrating the Sun, while lower energy showers develop at the solar depth where they are produced. Notice also that $E_{\rm c}$ is
a factor of 2 larger when the parent particle is a He nucleus. }
\beq
p_{\rm out}={1\over 2}\, e^{-\left( E/ E_{\rm c} \right)^2}.
\label{pout}
\eeq
Eq.~(\ref{pout}) is
then the second basic hypothesis in this framework, with the precise value of $E_c$ chosen to reproduce the drop in the 
gamma ray flux observed at $E>200$ GeV in Fig.~\ref{f4}.

Under the two assumptions expressed in Eqs.~(\ref{depth},\ref{pout}), we use cascade equations \cite{Gaisser:1990vg}
to find the final  flux of neutral particles (see also \cite{Masip:2017gvw,Gamez:2019dex}). 
The key difference with the usual showers in the Earth's atmosphere is due to the thin environment where these
solar showers develop (see Fig.~\ref{f1}): TeV pions and even muons decay there before they loose energy, defining a
neutrino flux well above the atmospheric one. In addition to this emission we must add the 
neutrinos produced in the back side of the Sun \cite{Edsjo:2017kjk,Ng:2017aur,Arguelles:2017eao}.
Our results for the signal in the different channels are the following.

In Fig.~\ref{f4}-left we plot the flux of gamma rays at $E>10$ GeV 
together with the Fermi-LAT ($E<200$ GeV) and HAWC ($E=1$ TeV) data. 
This spectrum  exhibits two main features. At low energies it is 
reduced because primary  CRs do not reach the Sun; notice that during a solar minimum the absorbed CR flux is larger, implying a more complete shadow and a larger gamma ray flux.
At higher energies the gamma flux is reduced as well, but for a different reason: 
all CRs reach the surface in their way to the Earth through 
the solar magnetic field and shower there, but most
photons are emitted towards the Sun and are absorbed. Although the set up does not provide a reason 
for the possible {\it dip} at 40 GeV \cite{Tang:2018wqp}, 
the 400--800 photons per squared meter and year that we obtain seem an acceptable fit to the data. We have not included leptonic gamma rays from inverse Compton scattering on solar photons 
\cite{Moskalenko:2006ta,Orlando:2006zs}. This contribution 
 extends beyond the solar disk \cite{Orlando:2020ezh} and is numerically important at $E<10$ GeV, 
possibly being also sizeable at $E>1$ TeV \cite{Zhou:2016ljf}.

An analogous flux of neutrons, most of them from the spallation of He nuclei in the solar surface, is given in Fig.~\ref{f4}-right. The plot shows the differential flux\footnote{Per sr, {\it i.e.}, the total neutron flux at the Earth is obtained multiplying by $\Omega_{\odot}\approx 6.5\times 10^{-5}$.} reaching the Earth averaged over  an 11 year solar cycle. Unfortunately, we have no hadronic calorimeters in space and this neutron flux (around 200 neutrons of energy above 10 GeV per squared meter and year) remains unobserved. 

The third line in Fig.~\ref{f4}-right corresponds to the average differential neutrino ($\nu+\bar \nu$)
flux from the solar surface. It includes both the albedo flux and the neutrinos produced in the back side that emerge
after crossing the Sun. The plot shows that at low energies neutrinos more than 
{\it doble} the number of gamma rays, whereas at $E>5$ TeV all albedo
fluxes vanish but we still get the fraction of neutrinos that are not absorbed after crossing the Sun.

\section{Neutrinos from  Jupiter and Earth} 

\begin{figure}[!t]
\begin{center}
\includegraphics[scale=0.48]{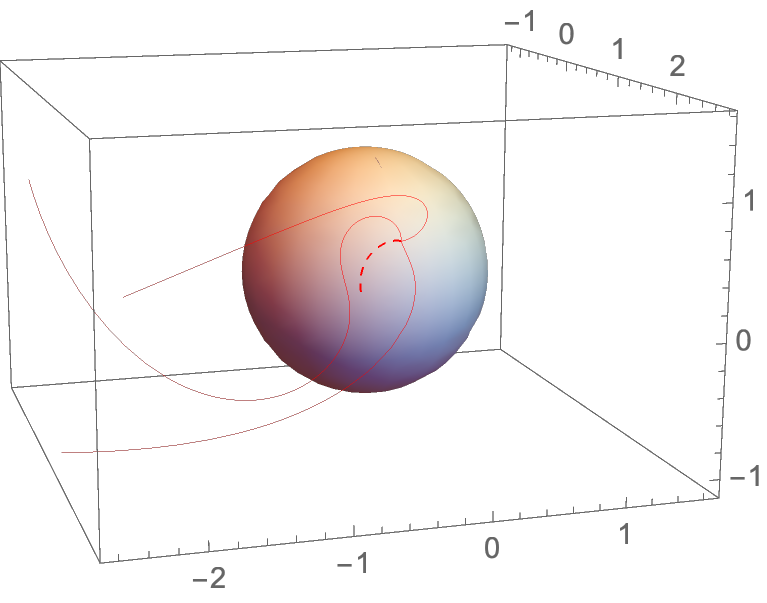}
\end{center}
\vspace{-0.3cm}
\caption{Trajectories of a 2 TeV proton reaching a point on the surface of Jupiter with the same zenith ($\theta_z=60^\circ$) and different azimuths (in units of $R_J$, with the center of the planet at $(0,0,0)$ and the
north pole at $(0,0,1)$).
The trajectory from the west (in dashes) is shadowed by Jupiter.
\label{f5}}
\end{figure}

Jupiter, like the Sun, has a thinner and more extended atmosphere than Earth (see Fig.~\ref{f1}). This will favor the decay of high energy mesons and muons in CR showers and thus the production of TeV neutrinos. Unlike the Sun, however, Jupiter does not emit a wind of plasma, and its magnetism is more similar to the one in Earth. As a consequence, the albedo flux of high energy gamma rays and neutrinos will  be negligible.

In our simplified model of Jupiter we will take a pure dipole magnetic field in a sphere of $R_J=6.99\times 10^4$ km, with the north  pole in the North geographic pole and $B_0=4.17\times 10^{-4}$ T 
at the equator on the surface \cite{Connerney:2017} (versus the opposite polarity and $B_0=3.12\times 10^{-5}$ T in  Earth).
This stronger magnetism implies an East-West effect (see Fig.~\ref{f5}) on higher energy CRs than in Earth: while in our planet low-latitude trajectories from the east are shadowed at  $E\lsim 15$ GeV, west trajectories in Jupiter are absent up to energies around 2 TeV.
\begin{figure}[!t]
\begin{center}
\hspace{-1.5cm}
\includegraphics[scale=0.48]{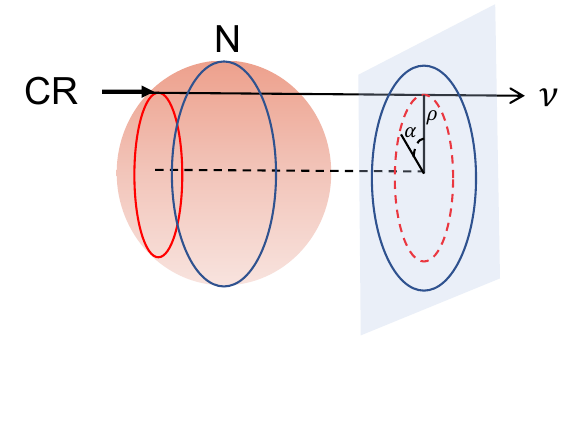}\hspace{0.5cm}
\includegraphics[scale=0.45]{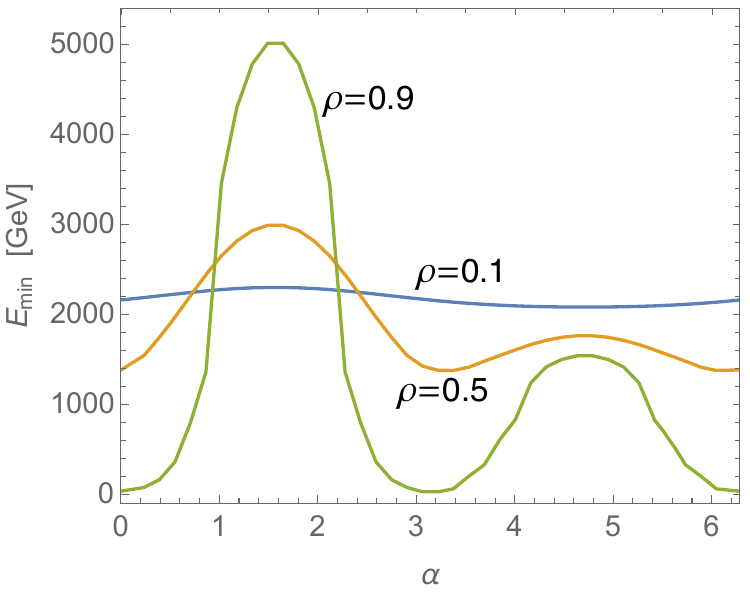} \hspace{5.5cm}$\,$\\
\vspace{-2.2cm}
$\;$ \hspace{10.5cm} \includegraphics[scale=0.48]{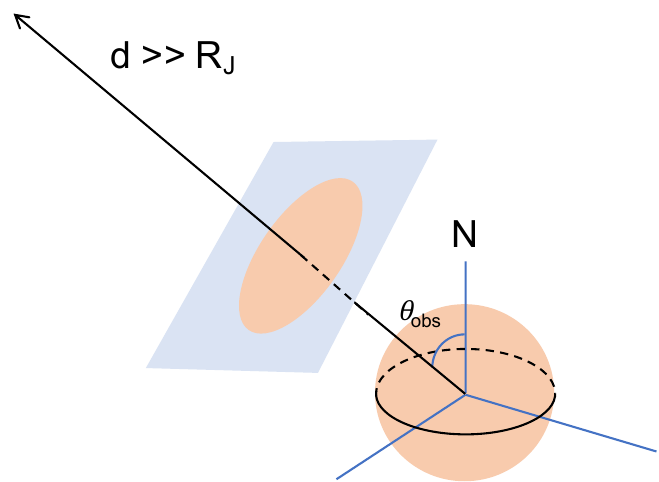} \\
\vspace{-1.6cm}
\hspace{-1.5cm}
\includegraphics[scale=0.48]{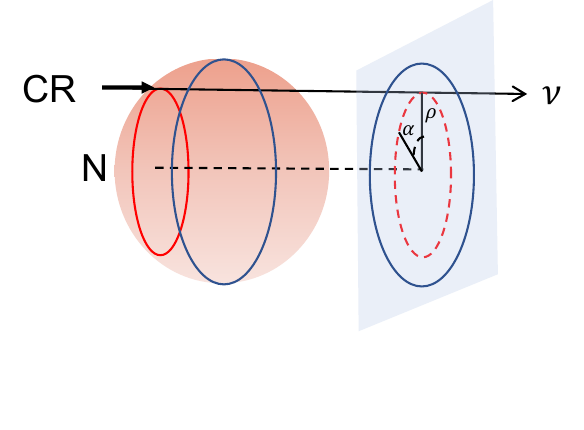}\hspace{0.5cm}
\includegraphics[scale=0.45]{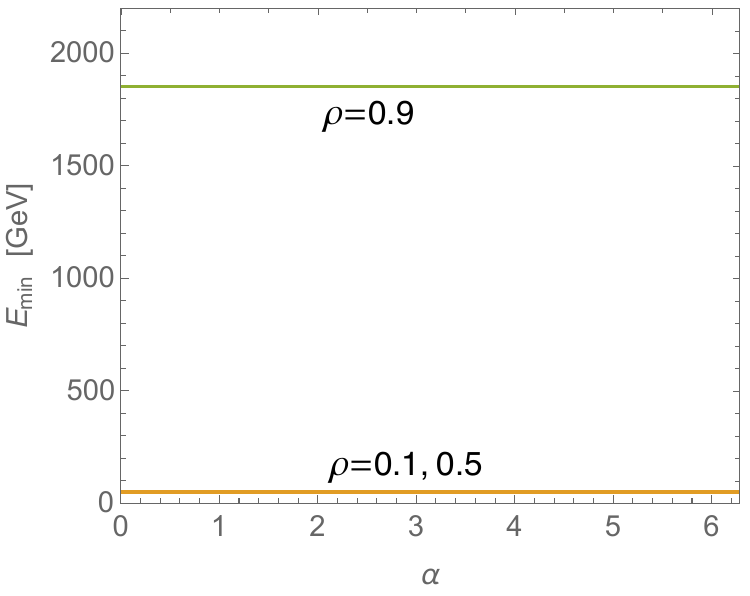}\hspace{5.6cm}$\,$
\end{center}
\vspace{-0.3cm}
\caption{Minimum energy of a CR proton able to reach the surface of Jupiter for different values of $\rho$ and $\alpha$ and for two relative orientations of the magnetic north pole. On the right, observation angle $\theta_{\rm obs}$; the position of the north pole in the disk goes from 
$\rho=0$ for $\theta_{\rm obs}=0$ to 
$\rho=1$ for $\theta_{\rm obs}=\pi/2$ and back to $\rho=0$ for $\theta_{\rm obs}=\pi$, always with $\alpha=0$.
\label{f6}}
\end{figure}

Let us first focus on Jupiter. In Fig.~\ref{f6} we provide the minimum CR energy ($E_{\rm min}$) that is not shadowed by magnetic effects when we observe the planet from a large distance ($d\gg R_J$) and two different directions. The radial parameter $0\le \rho\le 1$ and the angle $0\le \alpha <2\pi$ shown in the figure label all the points in the Jovian disk, with the position of the north pole at $\alpha=0$ from any observation direction. After a CR enters its atmosphere at a given point with a certain inclination, the neutrinos produced will cross Jupiter and emerge from the opposite side. 
The neutrino flux from the circle of constant $\rho$  will not be homogeneous, it changes with $\alpha$ as $E_{\rm min}$ varies at each point on the surface. The plot shows that if we look at Jupiter from the ecliptic plane (Fig.~\ref{f6}-top), lower energy CRs reach Jupiter only near the poles ($\alpha=0,\,\pi$), while West peripheral regions ($\alpha=\pi/2$) require a minimum CR energy of up to 5 TeV. 
Notice  that this minimum energy depends on the magnetic rigidity and will double when the primary is a He nucleus.

\begin{figure}[t!]
\begin{center}
\includegraphics[scale=0.49]{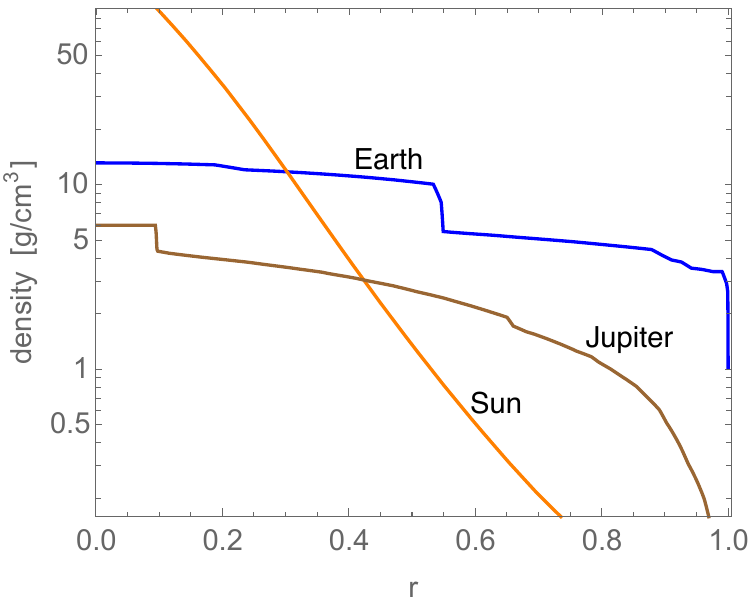}\hspace{1cm}
\includegraphics[scale=0.51]{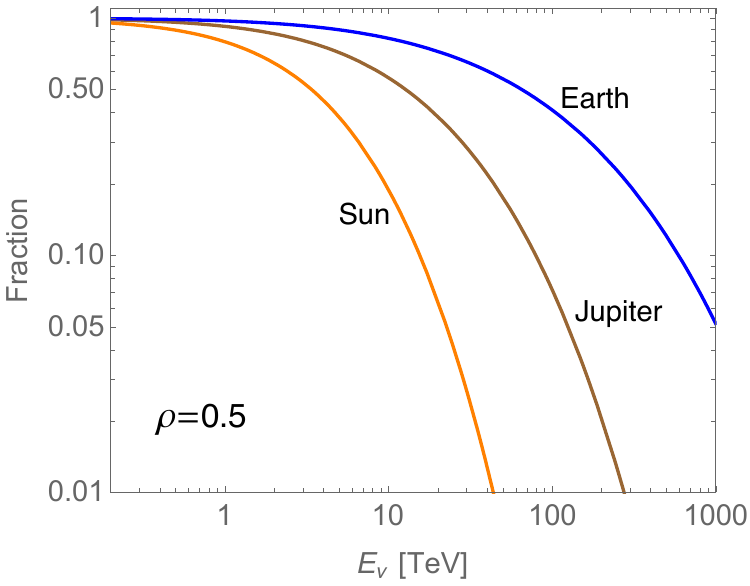}
\end{center}
\vspace{-0.3cm}
\caption{Density profile of the each astrophysical
object ($r$ is the normalized radius) and fraction of neutrinos not absorbed through a trajectory with $\rho=0.5$
(see Fig.~\ref{f6}).
\label{f7}}
\end{figure}

In addition, some of the neutrinos produced in the shower will be absorbed as they cross Jupiter, a possibility that depends on their energy and on the column density that they face before emerging (different for each value of 
$\rho$, see Fig.~\ref{f6}). 
In Fig.~\ref{f7}b we illustrate this absorption probability for a neutrino of different energy crossing the planet from 
$\rho=0.5$, including in the plot the analogous probability for the Sun and Earth. Due to the different density profiles
(given in Fig.~\ref{f7}a, we use \cite{Connerney:2017} for Jupiter and \cite{Dziewonski:1981xy} for Earth), the $\rho$ dependence 
of this  probability is much stronger for the Sun than for the two planets. For example,  
89\% of 500 GeV neutrinos can cross the Sun from $\rho=0.5$, but only 0.2\% from $\rho=0$; in contrast $72\%$ of 5 TeV neutrinos cross Jupiter from $\rho=0.5$ and $54\%$ from $\rho=0$. 
\begin{figure}[!b]
\begin{center}
\includegraphics[scale=0.48]{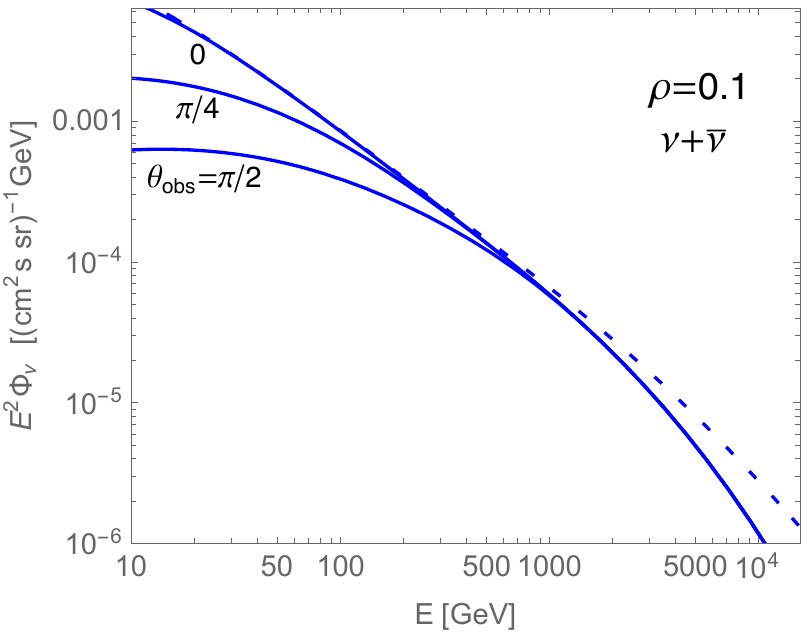}\hspace{1cm}
\includegraphics[scale=0.48]{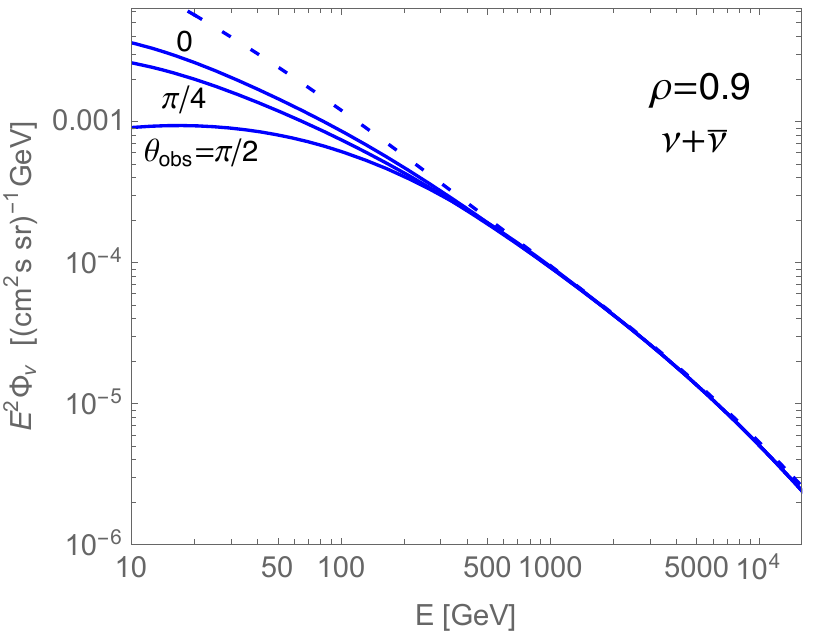}
\end{center}
\vspace{-0.3cm}
\caption{Neutrino flux from Jupiter observed from three directions $\theta_{\rm obs}$ at $\rho=0.1,\, 0.9$ (see Fig.~\ref{f6}). The dashed line is the flux before including magnetic and absorption effects.
\label{f9}}
\end{figure}

We can now find the neutrino flux when we observe Jupiter from a particular direction $\theta_{\rm obs}$ (see Fig.~\ref{f6}-right) from a distance $d\gg R_J$; in particular, we first
obtain the flux $\phi_\nu(E,\theta_{\rm obs},\rho)$ at $\rho$ 
averaged over $\alpha$.
For example, in Fig.~\ref{f9} we plot the neutrino flux when we observe Jupiter from three different angles 
($\theta_{\rm obs}=\pi/2,\,\pi/4,\,0$) at $\rho=0.1,\,0.9$. The dashed lines indicate
the flux after a slant depth of 7 km w.e. if we had ignored the Jovian magnetic field, whereas the solid lines include both the East-West effect (stronger when we observe from the ecliptic plane, $\theta_{\rm obs}=\pi/2$, than from $\theta_{\rm obs}=0$) and the absorption (that affects high energy neutrinos from $\rho=0.1$ but not the neutrinos from CRs skimming Jupiter at $\rho=0.9$).

Notice also that CR showers at $\rho=0.9$, being more {\it horizontal} than the ones at $\rho=0.1$, develop in a thinner environment and imply a larger neutrino flux. 

\begin{figure}[t!]
\begin{center}
\includegraphics[scale=0.5]{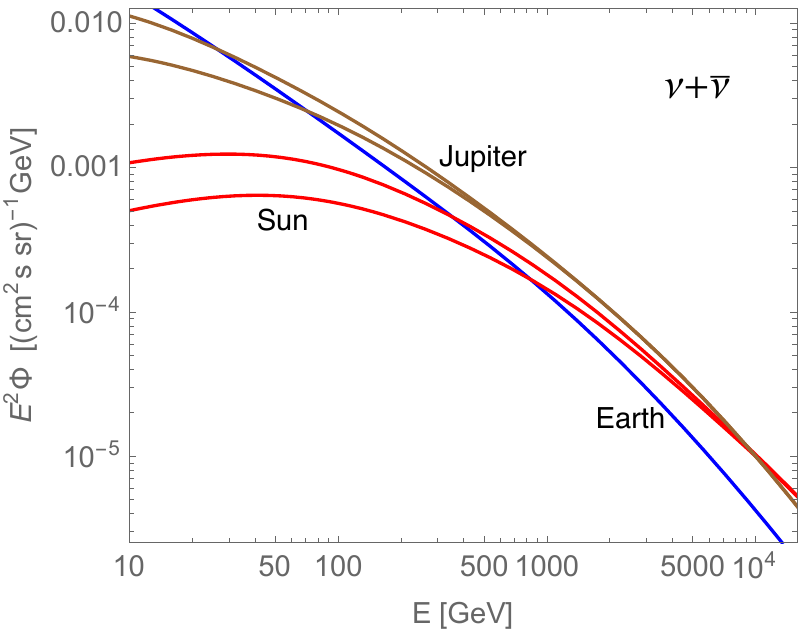}
\end{center}
\vspace{-0.3cm}
\caption{Neutrino flux averaged over the disks of the Sun, Jupiter and Earth  at $d\gg R_i$. 
\label{f10}}
\end{figure}

We finally obtain the flux $\phi_\nu(E,\theta_{\rm obs})$
averaged over the whole solid angle $\Delta \Omega_J$ occupied by the planet, {\it i.e.}, the total flux
(neutrinos per unit area, time and energy) is just $\phi_\nu\times \Delta \Omega_J$, where 
$\Delta \Omega_J$ depends on the distance $d$ of observation:
\beq
\Delta \Omega_J \approx {\pi \left( R_J/ d \right)^2\over 4\pi}\,.
\eeq
Fig.~\ref{f10} shows the average flux from the disk defined by Jupiter (from $\theta_{\rm obs}=\pi/2,\,0$), together with the analogous neutrino flux from the Sun and from Earth when observed from a large distance. 

The calculation of the neutrino flux emitted by
 Earth in Fig.~\ref{f10} 
has been analogous to the one done for Jupiter, with three basic differences. First, the East-West effect is opposite 
($\alpha\to 2\pi-\alpha$ in Fig.~\ref{f6}) and weaker than the one in Jupiter,
\beq
E_{\rm min}^E \approx {(B_0 R)_E\over  (B_0 R)_J} \,E_{\rm min}^J =  
6.8\times 10^{-3}\, E_{\rm min}^J\,.
\eeq
Second, while Jupiter has not a rocky surface, in Earth 
 CR showers develop only until they reach the ground, where all muons are absorbed 
before they can decay and produce neutrinos. And finally, the absorption of TeV neutrinos by Earth  is negligible, but 
not by Jupiter. 

These three features and also the different atmospheric densities define the neutrino fluxes shown  
in Fig.~\ref{f10},  where we have also included the one from the Sun during a solar max 
 (lower line) and a solar min (upper line) discussed in the previous section. 
At 10--500 GeV magnetic effects reduce the number of CRs reaching the surface of the Sun and of Jupiter but not of Earth, which implies a larger $\nu$ flux off the terrestrial disk at low energies.
At $500$ GeV the spectral index [see Eq.~(\ref{fluxp})] in these fluxes is, respectively, 2.7,  2.9 and
3.1. At $E> 500$ GeV
the flux from Earth becomes relatively smaller because showers develop there in a more dense
medium, which reduces the probability of  meson decays. 

The three fluxes obtained are much stronger than the diffuse background from CR interactions with the interstellar
gas. In particular, this flux averaged over the whole sky can be estimated
 [in ${\rm (GeV\,cm^2\,sr\,s)^{-1}}$] \cite{Carceller:2016upo}
\beq
\bar \Phi_\nu^{\rm gal} = 
3.7\times 10^{-6} \left({E\over {\rm GeV}}\right)^{-2.617}
+
0.9\times 10^{-6} \left({E\over {\rm GeV}}\right)^{-2.538}
\,,
\label{fluxnuG0}
\eeq
being a factor of 3 larger from the inner Galactic region ($|b|<8^\circ$, $|\ell |<80^\circ$) \cite{Schwefer:2022zly}. At
500 GeV, however, the neutrino flux from Jupiter is over 4000 times larger than the average diffuse background.

\section{Fluxes of gamma rays} 
Jupiter and Earth  emit high energy gamma rays produced by CRs that enter their atmosphere horizontally, skimming the planet from a zenith angle $\theta_z\approx 90^\circ$. Therefore, from space we will observe gamma rays coming only from the very peripheral region. We find, in particular,  that most gamma rays emerge from Jupiter at values of 
$\rho$ between 1.001 and 1.002. At lower values all the gamma rays are absorbed before they escape into space, whereas at larger values of $\rho$ the atmosphere is too thin and the production of gamma rays becomes negligible. 

\begin{figure}[t!]
\begin{center}
\includegraphics[scale=0.52]{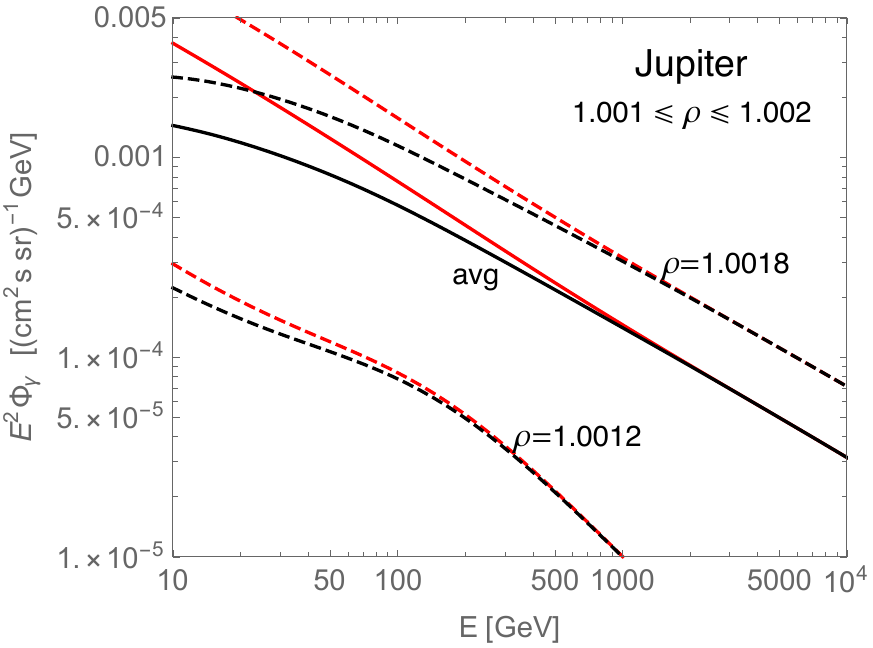}\hspace{1cm}
\includegraphics[scale=0.52]{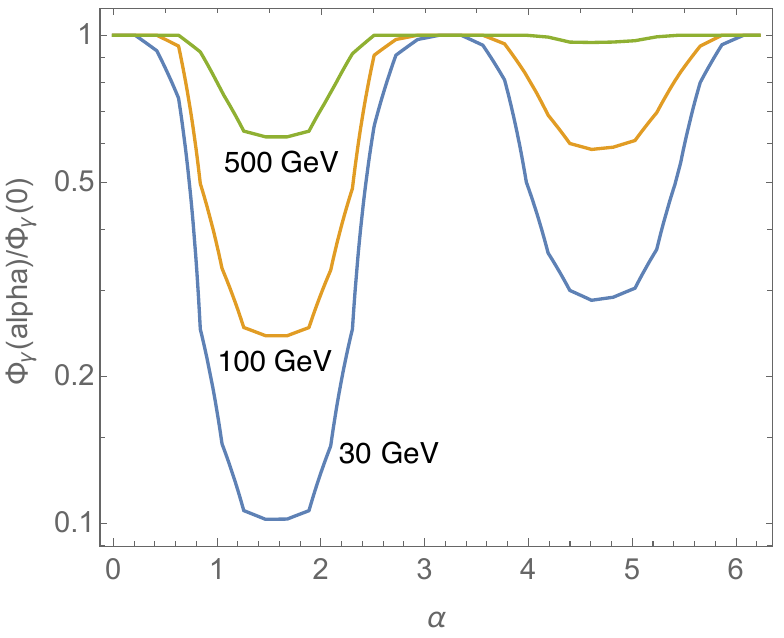}
\end{center}
\vspace{-0.3cm}
\caption{Gamma flux from the peripheral region of Jupiter (left) and East-West dependency (East at $\alpha= \pi/2$) 
of the flux.
\label{f11}}
\end{figure}

In Fig.~\ref{f11}--left we plot the gamma flux emitted by Jupiter for two particular values of $\rho$ together with the average flux from the annulus $1.001 \le \rho \le 1.002$ for an observation from the ecliptic plane ($\theta_{\rm obs}= \pi/2$, we neglect the tilt in the dipolar magnetic field). 
The red lines indicate the  flux if we had ignored the 
East-West effect, that is included in the final black lines. If the gamma ray detector is placed at a distance $d\approx 10R_J$, it implies around 2500 gamma rays of energy above
10 GeV per m$^2$ and year (for comparison, the neutrino flux from the whole Jupiter disk would be around 1600 times larger).
As discussed before, the minimum CR energy reaching the surface changes with $\alpha$ (see Fig.~\ref{f6}), which induces the dependence for the gamma ray flux given
in Fig.~\ref{f11}--right for three different energies. For example, the flux of 30 GeV gamma rays from the East side of the peripheral annulus
is  just a 10\% of the flux from the poles or 1/3 of the flux from the West. At 500 GeV the East-West effect becomes weaker, {\it e.g.},  the flux from the East is just reduced to a 60\% of the flux in the rest of the annulus.

\begin{figure}[t!]
\begin{center}
\includegraphics[scale=0.52]{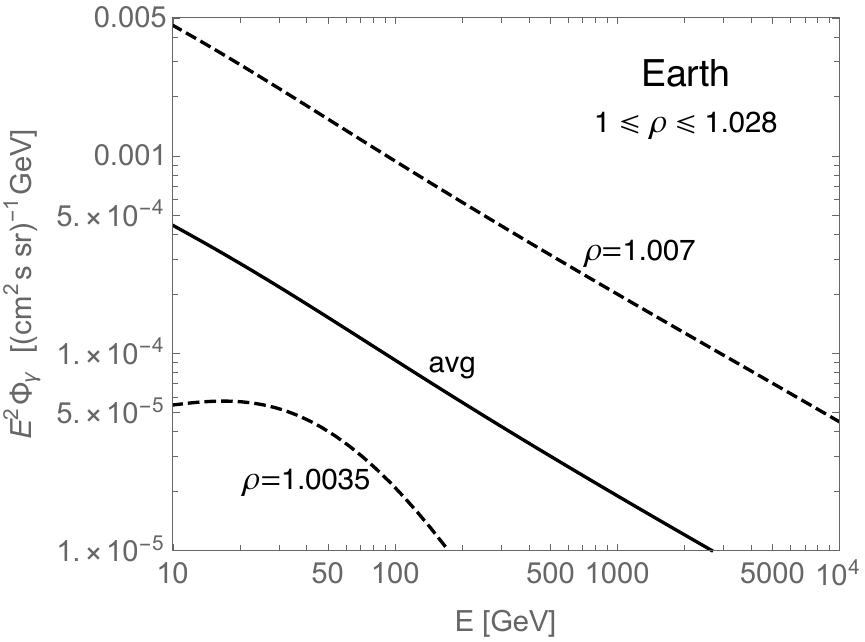}
\end{center}
\vspace{-0.3cm}
\caption{Gamma flux from the Earth. 
\label{f12}}
\end{figure}

Fig.~\ref{f12} gives the analogous gamma ray flux coming from Earth (as seen from space at $d\gg R_T$)), where the average refers to the $1\le \rho \le 1.028$ annulus (which corresponds to nadir angles between $66^\circ$ and
$70^\circ$ at Fermi-LAT, see below) and the East-West effect at $E_\gamma>10$ GeV is negligible (one expects an
 $\alpha$ dependence similar to the one in Fig.~\ref{f11}--right after the changes $\alpha\to 2\pi - \alpha$ and $E_\gamma\to 0.0068 \,E_\gamma $).
 The gamma ray flux from the Earth's limb has been investigated by FermiLAT in \cite{Fermi-LAT:2009dsu} and \cite{Madlee:2020ja} (in \cite{Fermi-LAT:2014qau} it is used to infer the CR spectrum). FermiLAT is located near the Earth ($d\not\gg R_T$) and their analysis focuses on lower energies, which are 
sensitive to the East-West effect and to the modulation of the primordial CR flux during the solar cycle. Their observations, however, are consistent with our results: at $E=100$ GeV and averaging the gamma ray flux over $66^\circ< \Theta_{\rm nadir} < 70^\circ$,
they find a spectral index $2.79\pm 0.06$ and 
${\rm d} N/{\rm d}  E= (8.9\pm 3.5)\times 10^{-9}$ (GeV cm$^2$ s sr)$^{-1}$, versus our values $2.71$ and $9.2\times 10^{-9}$ (GeV cm$^2$ s sr)$^{-1}$, respectively.

\section{Summary and discussion} 

Neutral astroparticles provide a picture of the sky complementary to the one obtained with light. It is then important
to identify and characterize all the known sources, as a necessary first step in the search for exotic phenomena.
In particular, indirect searches for dark matter target  
high energy gamma rays and neutrinos emitted from a variety of astrophysical objects, from the Sun to the Galactic center.

Here we have discussed the emission induced by CRs propagating near the surface of stars and  planets with a thin
atmosphere. In principle, one may expect that when high energy CRs reach the object all  secondary gamma rays except for the very peripheral ones will be absorbed, while a significant fraction of neutrinos will escape into space. This is indeed  the right picture at very high CR energies, when the magnetic deflection of charged particles can be neglected, but not at lower energies. 

Among the three objects considered, we find that the Sun is the most efficient source of gamma rays, specially at
100--1000 GeV. The key reason is that 1--50 TeV CRs are very affected by the magnetic field in the vicinity of the Sun, which induces an albedo  
including a large fraction (up to 50\%) of all secondary gamma rays. 
At lower energies most primary CRs are mirrored by the solar wind and do not reach the solar surface, 
whereas at larger energies
CRs do reach the Sun, but most secondary gamma rays are inwards and do not escape into space. 

In the 100--1000 GeV region the solar flux of gamma rays is roughly half the one of neutrinos (see Fig.~\ref{f4}), with a spectral index similar to the one in the parent CRs ($\sim E^{-2.7}$). In this energy interval the $\nu$ flux emitted by  Earth has a softer spectrum ($\sim E^{-3.7}$), which reflects the more dense environment where terrestrial showers develop.
At larger energies the absorption of neutrinos from the central regions ($\rho<0.5$) of the Sun (and also Jupiter, but not Earth) increases, and the spectral index softens. Fig.~\ref{f10} shows that the dipolar magnetic field in the two planets makes them a more efficient source of lower energy neutrinos ($E\le 500$ GeV), whereas at higher energies the determinant factor becomes the density in each object: Jupiter and the Sun have a thinner atmosphere than Earth and thus a larger $\nu$ flux.

In contrast to the flux from the Sun, the gamma flux from the two planets is restricted to the very peripheral region, implying almost two thousand times less gamma rays than  neutrinos. However, the East-West dependence of this
peripheral flux seems and interesting effect, as it provides information about the intensity and the structure of the planet's magnetic field. 

At $E_\gamma> 1$ TeV the solar albedo also disappears and the three sources  produce neutrinos but a much smaller flux of gamma rays. We find this remarkable as well, since it defines a very clear difference with the fluxes from CR interactions with the interstellar (and intergalactic) gas. At TeV--PeV energies these neutrino and gamma diffuse fluxes have a strong correlation, they have similar intensities and spectral indexes. IceCube has observed PeV neutrinos, and the extrapolation of this observed $\nu$ flux to lower energies is experimentally excluded if it comes together with a twin $\gamma$ flux \cite{Murase:2013rfa}. Our analysis suggests that astrophysical regions that are thin ($\rho\le 10^{-3}$ g/cm$^3$) but deep enough to absorb most gamma rays in the shower (a total depth larger $10^3$ g/cm$^2$) may play a role in the origin of the high energy 
IceCube \cite{IceCube:2013low} and KM3NeT \cite{KM3NeT:2025npi} events.

\section*{Acknowledgments}
This work has been supported by the Spanish Ministry of Science, Innovation and Universities 
MICIU/AEI/ 10.13039/501100011033/ (PID2022-14044NB-C21), by Junta de Andaluc{\'\i}a (FQM 101) and by 
Uni\'on Europea-NextGenerationEU ($\mathrm{AST22\_8.4}$).

\end{document}